\documentclass[a4paper,fleqn]{cas-sc}
\usepackage{placeins}

\usepackage[numbers]{natbib}

\def\tsc#1{\csdef{#1}{\textsc{\lowercase{#1}}\xspace}}
\tsc{WGM}
\tsc{QE}

\begin{document}
\let\WriteBookmarks\relax
\def\floatpagepagefraction{1}
\def\textpagefraction{.001}

\shorttitle{electrochemical tomography}    

\shortauthors{van Ede et al.}  

\title [mode = title]{Nondestructive detection and quantification of localized corrosion rates by electrochemical tomography}  



%

\author[1]{M.C. van Ede}

\credit{Conceptualization, Methodology, Validation, Investigation, Data Curation, Visualization, Writing - Original Draft}

\affiliation[1]{organization={Institute for Building Materials, ETH Zurich},
            city={Zurich},
            postcode={8093}, 
            country={Switzerland}}

\author[2]{A. Fichtner}

\credit{Conceptualization, Writing - Review \& Editing}

\affiliation[2]{organization={Institute of Geophysics, ETH Zurich},
            city={Zurich},
            postcode={8092}, 
            country={Switzerland}}

\author[1]{U. Angst}

\cormark[1]

\ead{uangst@ethz.ch}

\credit{Conceptualization, Writing - Review \& Editing, Supervision}

\cortext[1]{Corresponding author}


\begin{abstract}
Localized corrosion is one of the most common causes of early degradation of engineering structures. To non-destructively determine the location, size and rate of localized corrosion in porous media, a new technique, electrochemical tomography (ECT), has been theoretically and numerically formulated. The current work shows the application of ECT to measure corrosion rates in a controlled laboratory setup, with a stable electrolyte and well-defined macro-cell. The results show that ECT is able to replicate the corrosion size and location and can give a good estimation of the corrosion rate. Moreover, the validation of ECT on a well defined localized corrosion system, brings the technique closer to future field applications. 
\end{abstract}




\maketitle

\section{Introduction}
Non-destructive testing (NDT) techniques are becoming increasingly important in the condition assessment of civil infrastructure \cite{kurz2013condition, ahmed2020review}. The use of NDT enables detection of deterioration at an early stage, where it is often not yet visible during conventional visual inspection. Combined with advances made in automated inspection, e.g. with the use of robots, NDT techniques could significantly reduce the economic impact related to ageing infrastructure, which is a major challenge in all industrialized countries \cite{polder2012non, angst2018challenges}. Corrosion is one of the most common causes of early degradation of engineering structures, causing safety hazards and bringing along huge maintenance costs \cite{hansson2011impact}. In the United States, corrosion of infrastructure was estimated to cost more than 22 billion dollars annually \cite{koch2002corrosion}. In Switzerland, a recent study showed that more than half of the maintenance costs of road bridges is related to corrosion \cite{yilmaz2020korrosionsbedingte}. \\\\
Localized corrosion, or macro-cell corrosion, is the most dangerous type of corrosion, as large cathodic areas can result in high corrosion rates at small local anodes. This type of corrosion is especially problematic in porous media, such as reinforced concrete \cite{raupach1996chloride,elsener2002macrocell} and underground steel structures \cite{romanoff1957underground,gardiner2002corrosion, cole2012science}. To non-destructively detect localized corrosion in porous media, half-cell potential mapping is a well established and reliable technique \cite{elsener2003half,kessler2017reliability}, widely applied on-site, sometimes in the combination with electrical resistivity measurements \cite{polder2001test}. After the detection of areas with a high-risk of corrosion, additional destructive testing, such as drilling cores to determine chloride profiles, are used to assess if repair is needed \cite{glass2000chloride}. Not only are these destructive investigations costly, the measured chloride content cannot always reliably tell if corrosion has initiated, as the critical chloride content needed for initiation is influenced by a multitude of factors, such as the cement type \cite{angst2022beyond}. For the efficient scheduling of maintenance works, it would be more effective to directly non-destructively locate and quantify the localized corrosion.\\\\
However, as of now, there are no commercially available techniques that can reliably quantify the rate of localized corrosion non-destructively in the field \cite{nygaard2009corrosion}. Existing electrochemical techniques, that aim to measure corrosion rates, mostly rely on the measurement of the polarization resistance, e.g., by the linear polarisation resistance (LPR) technique \cite{nygaard2009corrosion, gowers1993site,mansfeld2009fundamental}, electrochemical impedance spectroscopy \cite{john1981use} and the galvanic pulse technique \cite{nygaard2009corrosion, elsner1995corrosion}. A large source of error arises from the assumption of uniform corrosion and the need to estimate the polarized area of the steel \cite{song2000theoretical}. Therefore, these techniques are often found to yield unreliable estimations \cite{law2004measurement,nygaard2009non} and their applicability for localized corrosion has been questioned \cite{song2000theoretical,clement2012numerical, angst2015applicability}.\\\\
These limitations may be overcome by the combined use of a numerical model, simulating the potential field in the porous medium around the steel, and an inverse method. This idea has been formulated theoretically and numerically in previous work, and has been named electrochemical tomography (ECT) \cite{van2021electrochemical}. Using electrical potential values, measured at the surface of the porous medium for small externally applied currents, ECT determines the location and size of the corroding steel, as well as the corrosion rate. As ECT explicitly takes into account the geometry, electrical resistivity and the polarization behaviour of the steel, ECT can be applied to quantify localized corrosion and is not limited by the need for assumptions related to the (unknown) polarized area. \\\\
In van Ede et al. (2021) \cite{van2021electrochemical}, the authors showed that the technique is able to accurately estimate the corrosion location and size, as well as the corrosion rate, and give insight about the uncertainty associated with the solution. Moreover, this work also showed that the technique is able to study fundamental parameters that describe the corrosion kinetics of the steel in a macro-cell corrosion system, such as the Tafel slopes and exchange current densities.\\\\
Others have also shown that the use of inverse techniques is promising to increase the information obtained by potential mapping \cite{aoki1998elimination,ridha2001multistep,kranc2005numerical, marinier2013refined, adriman2022improving}. Marinier \& Isgor (2013) \cite{marinier2013refined} were the first to implement the Butler-Volmer kinetics, describing the polarization behaviour of the steel, into their inverse technique. Recently, Adriman et al. (2021) \cite{adriman2022improving} applied their deterministic inverse technique to the case study of a concrete pillar. They applied a more simplified approach than used in ECT, concerning the assumed corrosion kinetics and the required input data. Although their approach was not able to quantify the corrosion rate, they showed it was possible to accurately estimate the location and size of the localized corrosion area. \\\\
The current work further explores the technique of ECT and shows its first validation in a controlled laboratory setting. This validation is essential to evaluate the ability of the numerical model to correctly describe macro-cell corrosion, and to assess the accuracy of ECT in determining the corrosion location, size and rate in a real macro-cell corrosion setup. This work also further optimizes the technique to increase its accuracy and efficiency, in terms of the applied external currents during the measurement of the input potential data, and the corrosion kinetics fitted during the inversion procedure.

\section{Material and methods}
The goal of the validation of ECT \cite{van2021electrochemical} is to test if the technique as a whole, and the numerical model in particular, is accurate enough to estimate the location, the size and the rate of localized corrosion in a well defined geometry with a homogenous electrical resistivity. ECT was developed for locating and quantifying localized corrosion in porous media. However, direct validation in media such as soil or concrete is not ideal, as the steel surface, and thus the location and size of the corroding steel, cannot be observed during the experiments, and it is difficult to monitor and control the conditions of the electrolyte.  \\\\
Therefore we use a flow-cell with a pH neutral NaCl solution. The transparent flow-cell allows us to observe the steel during the experimental runs. Moreover, thanks to the continuous flow, the electrolyte around the steel surface can be maintained constant, as the solution stays well aerated and mixed during the experiment, and corrosion products are transported away from the steel surface. The macro-cell corrosion is simulated by using a steel bar consisting of two metals: carbon steel representing the corroding spot (anode), and stainless steel representing the surrounding passive steel (cathode). The galvanic interaction of these steels ensures that we have a well defined macro-cell, where the anode size and location are known, and the macro-cell current can be monitored by measuring the current flowing through the steel between the anode and cathode.\\\\
In the following we describe the experimental setup, the numerical model and settings of the ECT technique, and finally the experimental procedure.
\begin{figure}[h]
	\centering
	\includegraphics[width=\textwidth]{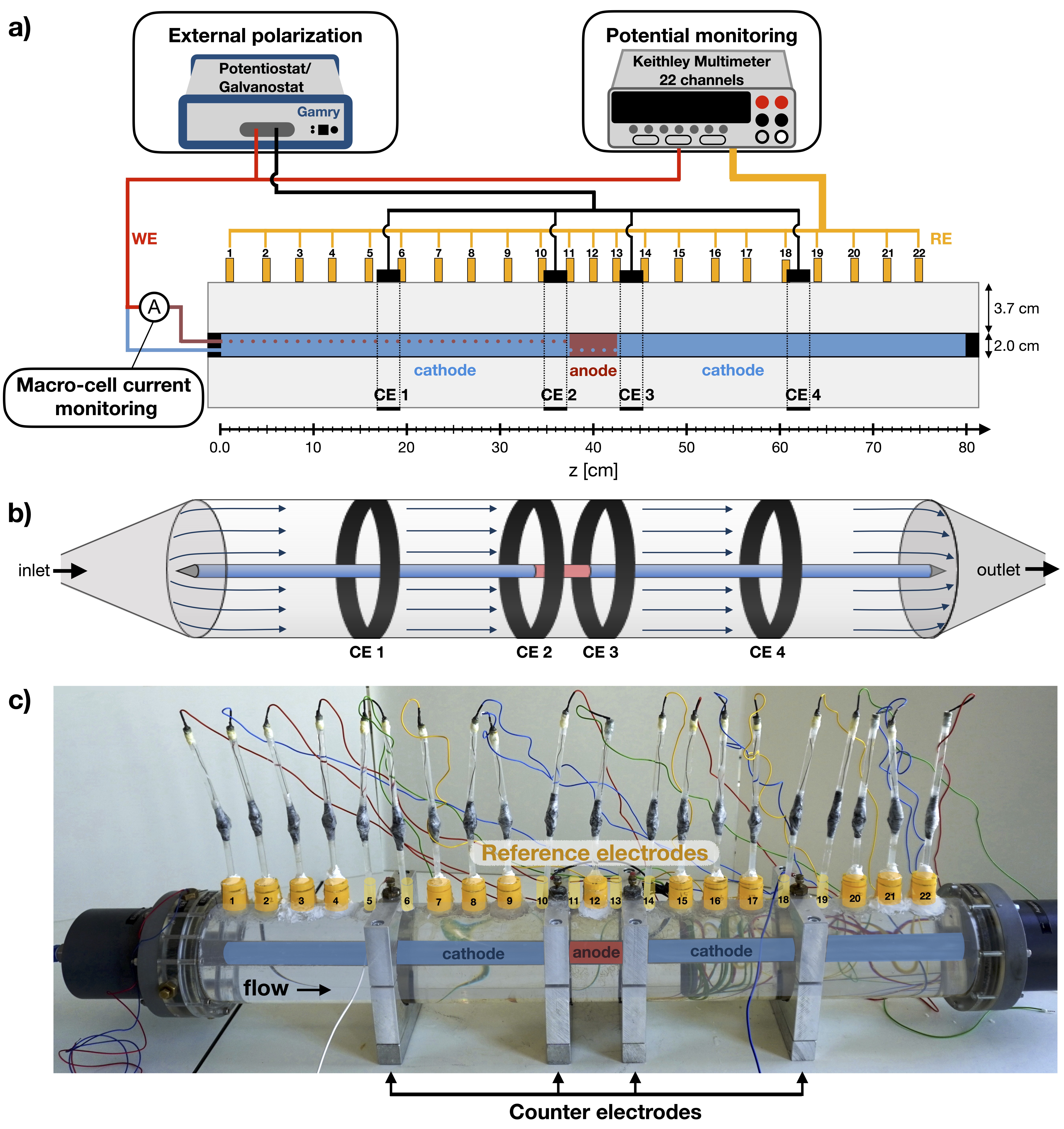}
	\caption{The experimental setup for the validation of ECT. a) Schematic overview showing the dimensions and the wire configurations to the working electrode (WE), the counter electrodes (CE) and the reference electrodes (RE). b) 3-dimensional overview of the flow-cell, ensuring quasi-laminar flow. c) Photograph of the laboratory setup, showing the flow-cell and all electrodes.}
	\label{fig:exp_overview}
\end{figure}
\FloatBarrier
\subsection{Experimental setup} \label{sec:exp_setup}
Figure \ref{fig:exp_overview}a gives an schematic overview of the laboratory setup, indicating the positions, dimensions and connections of the electrodes. A hollow cylindrical, smooth steel bar with a diameter  of 2 cm was placed in the middle of the cylindrical flow-cell, consisting of an SAE J2340 carbon steel (anode) and an X5CrNi18-10 stainless steel (cathode). The carbon steel was located in the centre of the flow-cell and had a length of 5 cm. The carbon and stainless steels were electrically connected with wire, that were guided through the hollow bar, outside of the flow-cell. This allowed us to measure the macro-cell current flowing between the anode and cathode with a zero resistance Ammeter (ZRA).\\\\
Four high grade stainless steel rings, with a radius of 4.7 cm and width of 2 cm, were placed at the outer rim of the flow-cell and acted as counter electrodes (CE). Two (CE1 \& CE4) were placed in the region of the cathode, and two (CE2 \& CE3) in the proximity of the anode. The CE, as well as the steel bar (working electrode WE), were connected to a Gamry Potentiostat/Galvanostat, in a 2-electrode setup. This setup was used to apply the external currents and polarize the system. Depending on the measurements, only CE1 \& CE4, only CE2 \& CE3, or all CE were connected (see section \ref{sec:procedure}).\\\\
At the surface of the flow-cell, 22 Ag/AgCl/Sat.KCl reference electrodes (RE) could be positioned. The RE were connected to a Keithley multimeter with 22 channels, which was used to monitor the electrical potential between the steel bar (WE) and each reference electrode, with 0.1 s intervals. \\\\
Figure \ref{fig:exp_overview}b shows the 3-dimensional shape of the flow-cell and figure \ref{fig:exp_overview}c the flow-cell in the laboratory setup, highlighting the steel bar, the CE, the RE and the flow direction. The flow-cell was connected to a 10 l reservoir, which was stirred and bubbled with air to constantly homogenize the electrolyte. From the reservoir, the solution was pumped to the flow-cell using a peristaltic pump, with a flow speed of around 500 ml/min, resulting in the solution in the flow-cell to be completely replaced around every 12 min. The solution entered and exited the flow-cell through cones (figure \ref{fig:exp_overview}b), minimizing the turbulence of the flow and leading to a quasi-laminar flow along the steel surface. From the other end of the flow-cell, the solution flowed back into the reservoir. We used a $\sim$0.46 mM NaCl solution, pH 7, with an electrical resistivity of 130-140 $\Omega$m.\\\\
The RE consisted of an AgCl wire in a glass capillary filled with saturated KCl solution. At the junction, the capillary was placed in a second capillary containing the flow-cell NaCl solution, to protect both the RE and the solution in the flow-cell: The KCl solution was hindered from leaking directly into the flow-cell, and the KCl in the RE had minimal influence on the flow-cell solution, ensuring the stability of the RE. These reference electrodes could be screwed on top of the flow-cell (see figure \ref{fig:exp_overview}c), so that they could be calibrated before each experiment against an external Ag/AgCl/Sat. KCL Reference electrode. RE 5, 6, 10, 11, 13, 14, 18 and 19 were permanently fixed on the flow-cell for a duration of 4 days maximum, due to space limitations. These electrodes were calibrated directly in the flow-cell before the measurements, placing the external reference electrode in the RE positions closest to the fixed RE.  

\begin{figure}[h]
	\centering
	\includegraphics[width=\textwidth]{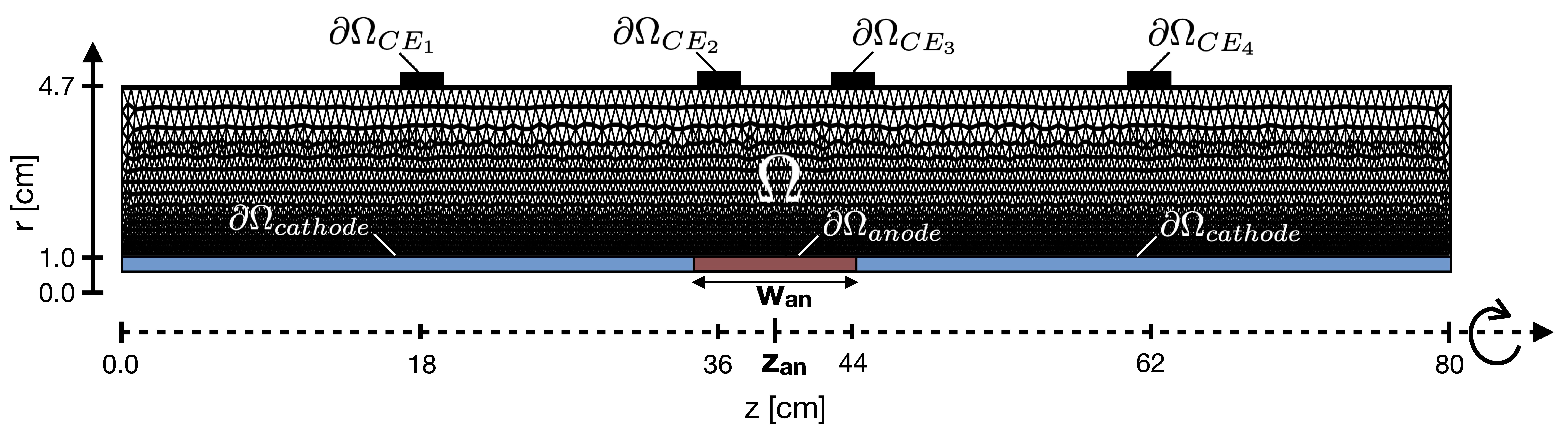}
	\caption{Overview of the geometry and mesh of the numerical model. The model is 2D axi-symmetric around the z-axis. Surfaces on which specific boundary conditions apply are labelled: the anodic and cathodic surface, and the counter electrodes (CE). The model parameters $W_{an}$, the width of the anode, and $z_{an}$, the location of the middle of the anode, are indicated.}
	\label{fig:num_model}
\end{figure}
\FloatBarrier

\subsection{ECT} \label{sec:meth_ECT}

ECT is described in detail in van Ede et al. (2021) \cite{van2021electrochemical}. It uses a stochastic inverse technique to find probability distributions for a set of model parameters that produce a good fit between the measured potentials at the surface and a numerical model. This numerical model simulates the electrical potential field, and thus the potential at the surface, for the investigated geometry. Figure \ref{fig:num_model} shows a schematic overview of the geometry and the mesh of the 2D axi-symmetric numerical model, used to represent the laboratory setup.  The mesh consists of tetrahedral elements, that have a maximum size of 5 mm at the outer surface and get finer towards the steel surface, where they approach a size of <1 mm. \\\\
For carbon steel, we assume that the main anodic reaction is the oxidation of iron. The main cathodic reaction for the stainless steel is the oxygen reduction reaction (ORR), as the solution is well aerated and around neutral pH. The strong-form formulations for the numerical model are derived in van Ede et al. (2021) \cite{van2021electrochemical} and are given by:
\begin{equation}
	\nabla^{2} \phi = 0 \ \  \text{in the electrolyte, }  \Omega
	\label{eq:main_gen}
\end{equation}

with  boundary conditions:

\begin{flalign}
	&\frac{d\phi}{d\overrightarrow{n}} = -\rho*i_{Fe,0} * \exp{\frac{2.303(\phi-\phi_{Fe}^{rev})}{\beta_{Fe, an}}}  \ \  \text{at the anode, }\partial\Omega_{anode}\text{,} \label{eq:bound_anode} &&  \\ 
	& \frac{d\phi}{d\overrightarrow{n}}= \rho*\left[ \frac{\left( i_{O_{2},0}\exp{\frac{2.303(\phi_{O_{2}}^{rev}-\phi)}{\beta_{O_{2}, cath}}}\right)^\gamma}{1+\left(\frac{i_{O_{2}, 0}}{i_{L}}\exp{\frac{2.303(\phi_{O_{2}}^{rev}-\phi)}{\beta_{O_{2}, cath}}}\right)^\gamma}              \right]^{\frac{1}{\gamma}}  \ \  \text{at the cathode, }\partial\Omega_{cathode}\text{,} \label{eq:inv_ox_gen}  &&\\
	&\frac{d\phi}{d\overrightarrow{n}} = -\rho * i_{CE} \ \  \text{at the CE, } \partial\Omega_{CE}\text{,}  &&\\
	&\frac{d\phi}{d\overrightarrow{n}} = 0 \ \ \text{at all other boundaries}\text{.}  \label{eq:other_boundaries}
\end{flalign}
Here, $\phi$ is the electrical potential, $\rho$ the electrical resistivity and $\overrightarrow{n}$ the normal to the surface. The reversible potentials, $\phi_{Fe}^{rev}$ and $\phi_{O_{2}}^{rev}$, can be computed using the Nernst equation \cite{mccafferty2010introduction}. The Tafel slopes of iron oxidation and oxygen reduction are given by $\beta_{Fe, an}$ and $\beta_{O_{2}, cath}$ respectively, and $i_{Fe,0}$ and $i_{O_{2},0}$ are the exchange current densities. The oxygen concentration controlled plateau of the oxygen reduction reaction is described by the limiting current density, $i_{L}$, which is, due to the aeration of the used solution, assumed to be high in the current setup (2 A/m$^2$) and $\gamma$ is a curvature defining constant, controlling the sharpness of the cathodic polarization curve in the transition from the activation-controlled to the concentration-controlled plateau \cite{kim2010modeling}. Finally, $i_{CE}$ is the current density at the counter electrodes, which is obtained by dividing the applied current by the total surface of the counter electrodes.\\\\
The numerical model simulates the potential field as a function of 6 model parameters: The width of the anode, $W_{an}$, and the location (z-coordinate) of the centre of the anode, $z_{an}$ (figure \ref{fig:num_model}) and four kinetic parameters, the Tafel slopes, $\beta_{Fe, an}$ and $\beta_{O_{2}, cath}$, and the exchange current densities, $i_{Fe,0}$ and $i_{O_{2},0}$. All other parameters are fixed and given in table \ref{tab:param}. \\\\
ECT uses a Bayesian inverse approach to obtain the posterior probability distribution of the model parameters, given the measured potentials at the surface and a-priori information about the model parameters \cite{van2021electrochemical}. The noise associated with the error of the potential measurements was assumed to be a Gaussian, with a variance of 0.005 V$^2$, which corresponds to a standard deviation of around 7 mV. The considered prior probability distributions for each model parameters are given in table \ref{tab:param}. As described in van Ede et al. (2021) \cite{van2021electrochemical}, the prior probabilities for $W_{an}$ and $z_{an}$ are determined from the shape of the measured potential distribution at the surface, when the steel is not polarized (see figure \ref{fig:repeatability}a). The priors for the kinetic parameters are based on experiments on stainless and carbon steel in an RDE setup at pH 7.5. In these experiments environmental factors, such as the convection of the solution and the condition of the steel surface (clean, corroded and containing a passive film), was varied, to obtain a realistic spread of the kinetic parameters that could be observed in a macro-cell setting (as presented for the ORR kinetics in \cite{vanEdekinetics}). The exchange current densities were assigned a Gaussian prior probability, while the Tafel slopes had a flat prior probability distribution, meaning that all values within the indicated range in table \ref{tab:param} are equally likely.\\\\
The marginal posterior distributions were sampled using a Markov chain Monte Carlo (MCMC) algorithm \cite{van2021electrochemical}. These chains contained 60,000 samples, obtained after 5,000 burn-in samples. The samples approximate the probability distribution for each model parameter separately. Finally the probability distribution for the corrosion rate was sampled, by integrating the current density at the anode surface, as described in van Ede et al. (2021) \cite{van2021electrochemical}.

\begin{table}[h]
	\renewcommand{\arraystretch}{1.2}
	\caption{The model constants fixed in the numerical model and the model parameters, considered in the inversion, with their respective prior distributions. 'Flat' refers to a constant probability distribution within the indicated range, 'Gaussian' to a normal distribution with an indicated mean, $\mu$, and standard deviation, $\sigma$}
	\label{tab:param}
	\begin{tabular}{lcl}
		\multicolumn{3}{c}{\textbf{Model constants}}                                                                                                                                                                \\
		\multicolumn{1}{l|}{\textbf{Parameter}}  & \textbf{Description}   & \multicolumn{1}{|c}{\textbf{Value {[}unit{]}}}                                                                                          \\ \hline
		\multicolumn{1}{l|}{$\phi_{Fe}^{rev}$}            & Reversible potential Fe oxidation (pH 7 \cite{pourbaix1974atlas}     & \multicolumn{1}{|c}{-0.61 {[}V vs SHE{]}}                                                                           \\
		\multicolumn{1}{l|}{$\phi_{O_{2}}^{rev}$}      & Reversible potential O$_{2}$ reduction (pH 7) \cite{pourbaix1974atlas}      & \multicolumn{1}{|c}{0.82 {[}V vs SHE{]}}                                                                                 \\
		\multicolumn{1}{l|}{$i_L$}              & Limiting current density O$_{2}$ (well aerated)               & \multicolumn{1}{|c}{2.0 {[}A/m2{]}}                                                                                \\
		\multicolumn{1}{l|}{$\gamma$}            & Curvature defining constant \cite{kim2010modeling}              & \multicolumn{1}{|c}{3}                                                                                                  \\ \hline
		\multicolumn{3}{c}{\textbf{Model parameters (solved by the inverse method)}}                                                                                                                                                               \\
		\multicolumn{1}{l|}{\textbf{Parameter}} & \textbf{Description}                 & \multicolumn{1}{|c}{\textbf{Prior {[}unit{]}}}                                                                             \\ \hline
		\multicolumn{1}{l|}{$W_{an}$}                     & Width of the anode           & \multicolumn{1}{|c}{\begin{tabular}[c]{@{}c@{}}'Flat' {[}m{]}\\ 0.01-0.3\end{tabular}}                                        \\
		\multicolumn{1}{l|}{$z_{an}$}           & Location of the middle of the anode         & \multicolumn{1}{|c}{\begin{tabular}[c]{@{}c@{}}'Gaussian' {[}m{]}\\ $\mu$ = 0.40, $\sigma$ = 0.075\end{tabular}}               \\
		\multicolumn{1}{l|}{$i_{Fe,0}$}        & Exchange current density Fe oxidation        & \multicolumn{1}{|c}{\begin{tabular}[c]{@{}c@{}}'Gaussian' log10({[}A/m$^2${]})\\ $\mu$ = -3.25, $\sigma$ = 0.880\end{tabular}}   \\
		\multicolumn{1}{l|}{$i_{O_{2},0}$}       & Exchange current density O$_{2}$ reduction        & \multicolumn{1}{|c}{\begin{tabular}[c]{@{}c@{}}'Gaussian' log10({[}A/m$^2${]})\\ $\mu$ = -6.98, $\sigma$ = 0.957\end{tabular}}  \\
		\multicolumn{1}{l|}{$\beta_{Fe, an}$}           & Anodic Tafel slope Fe oxidation     & \multicolumn{1}{|c}{\begin{tabular}[c]{@{}c@{}}'Flat' {[}V/dec{]}\\ 0.01-0.2\end{tabular}}                                    \\
		\multicolumn{1}{l|}{$\beta_{O_{2}, cath}$}        & Cathodic Tafel slope O$_{2}$ reduction           & \multicolumn{1}{|c}{\begin{tabular}[c]{@{}c@{}}'Flat' {[}V/dec{]}\\ 0.01-0.3\end{tabular}}                            
	\end{tabular}
\end{table}

\subsection{Experimental procedure} \label{sec:procedure}

To ensure the reliability of our validation data, we performed four independent experimental runs, where potentials were monitored for different externally applied currents at the CE. These 4 runs were performed in freshly prepared electrolytes, that had similar electrical resistivity, around 130-140 $\Omega$m  (table \ref{tab:runs}). Before each experiment, the steel bar was ground and degreased by hand, using grinding paper and ethanol, to remove corrosion products and ensure similar starting conditions for each experimental run. The 22 RE were calibrated against an external Ag/AgCl/Sat. KCl reference electrode, before placing them on the flow-cell. The RE that were fixed on the flow-cell (see section \ref{sec:exp_setup}), were calibrated directly in the flow-cell for run 1, run 2 and run 3. The NaCl solution was stirred and bubbled with air, before filling the flow-cell. As soon as the flow-cell was filled and the reference electrodes were placed, the flow was started and the open circuit electrical potentials (OCP), as well as the current between the anode and cathode were monitored.

\begin{table}[h!]
	\renewcommand{\arraystretch}{1.2}
	\centering
	\caption{ The performed repetitions of the experimental validation, with slightly varying resistivity of the electrolyte and submerge time of the steel bar in the flow-cell, before the start of the ECT measurement procedure.}
	\label{tab:runs}
		\begin{tabular}{l|c|c|c}
			\textbf{Exp.} & \textbf{\begin{tabular}[c]{@{}c@{}}Electrical resistivity \\ $\rho$ {[}$\Omega $m{]}\end{tabular}} & \textbf{\begin{tabular}[c]{@{}c@{}}submerge time \\ {[}hours{]}\end{tabular}} & \textbf{\begin{tabular}[c]{@{}c@{}}CE configurations\\ \end{tabular}}                                                                  \\ \hline
			run 1               &  140 +/- 2                                          & 16                                                                            &  'All CE' and 'CE2 \& CE3'  \\ 
			run 2               & 138 +/- 2                                                                                            & 17                                                                            &    'All CE' and 'CE2 \& CE3'                                                                                                                                                                                                                                   \\ 
			run 3               & 133 +/- 2                                                                                             & 16                                                                            &    'All CE', 'CE2 \& CE3' and 'CE1                                                                                                                                                                                                                          \\ 
			run 4               & 128 +/- 2                                                                                        & 20                                                                            &    'All CE' and'CE2 \& CE3'                                                                                                                                                                                                                                
		\end{tabular}
\end{table}

\begin{table}[h!]
	\renewcommand{\arraystretch}{1.2}
	\small
	\centering
	\caption{ The applied currents and the duration of the pulse (equal to the pause after the pulse) of the Galvanostatic pulse procedure (figure \ref{fig:flowchart})}
	\label{tab:applied_currents}
	\begin{tabular}{c|c}
		\textbf{\begin{tabular}[c]{@{}c@{}}Applied currents\\ {[}mA{]}\end{tabular}}                                                           & \textbf{\begin{tabular}[c]{@{}c@{}}Galvanostatic pulse duration\\ {[}min{]}\end{tabular}}          \\ \hline
		+ 0.10 & 5\\ 
		- 0.10 & 5\\ 
		+ 0.15 & 8\\ 
		- 0.15 & 8\\ 
		+ 0.20 & 10\\
		- 0.20 & 10\\ 
		+ 0.30 & 10\\ 
		+ 0.50 & 10\\ 
		+ 0.70 & 10 
	\end{tabular}
\end{table}

\begin{figure}[h]
	\centering
	\includegraphics[width=0.8\textwidth]{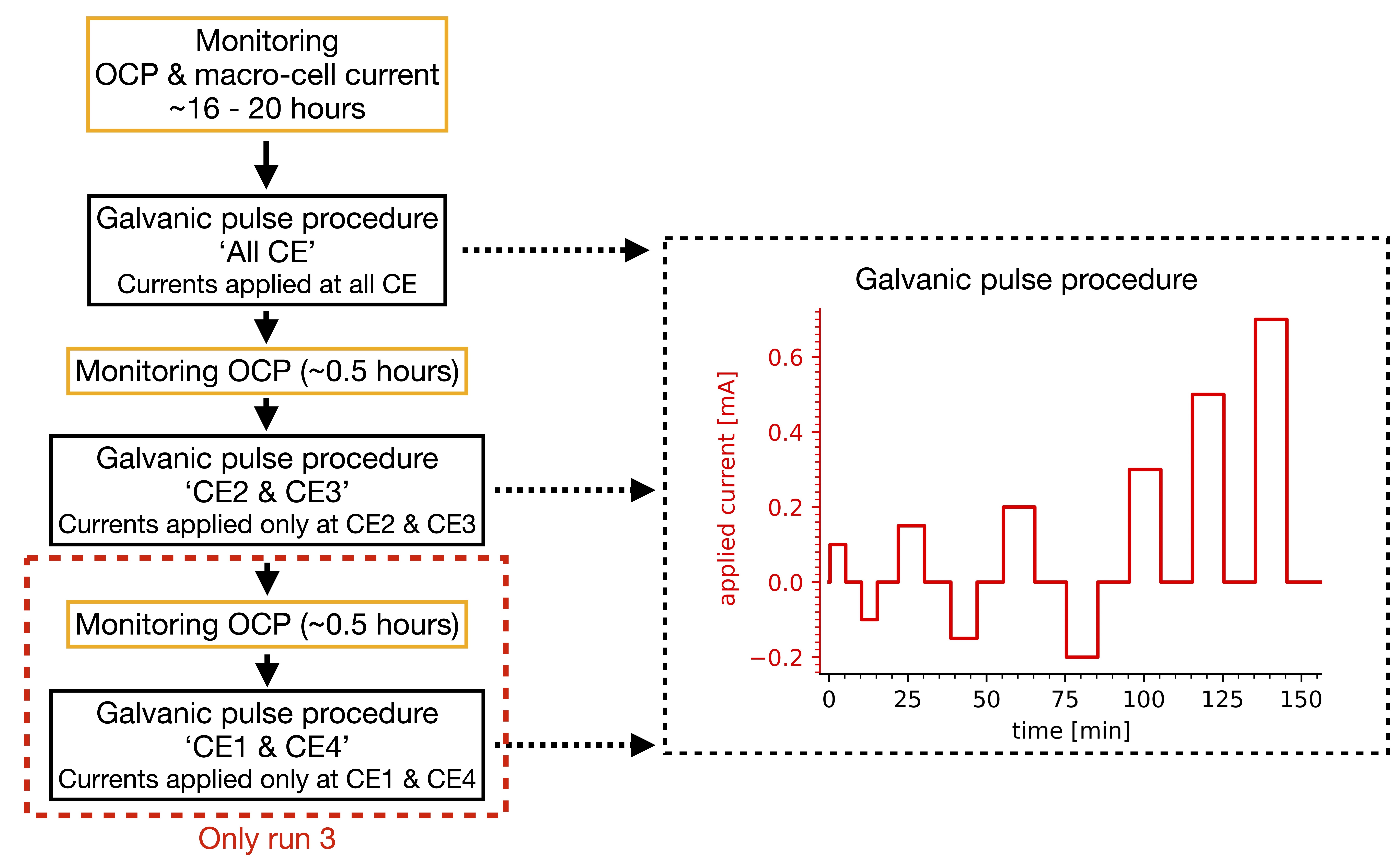}
	\caption{Flowchart of the ECT measurement procedure.}
	\label{fig:flowchart}
\end{figure}

After a submerge time of around 16-20 hours (table \ref{tab:runs}), the monitored electrical potentials and current reached stable values (see figure A1 in the supplementary materials) and the ECT measurement procedure was started. The complete procedure is summarized in the flowchart in figure \ref{fig:flowchart}. For different CE configurations ('All CE', 'CE2 \& CE3' and 'CE1 \& CE4'), 9 increasing cathodic and anodic currents were applied alternately for a certain polarization duration (table \ref{tab:applied_currents} and visualized in figure \ref{fig:flowchart}). After each individual applied current, the OCP was monitored for the same length of time as the polarization duration, before applying the next current step. Only in run 3 all three CE configurations were measured, due to time restrictions (table \ref{tab:runs}). During the experiments the resistivity of the solution was measured several times and was found to be stable within a +/- 2 $\Omega$m deviation. 

\FloatBarrier

\section{Results and Discussion}
\subsection{ECT validation}

\begin{figure}[h]
	\centering
	\includegraphics[width=\textwidth]{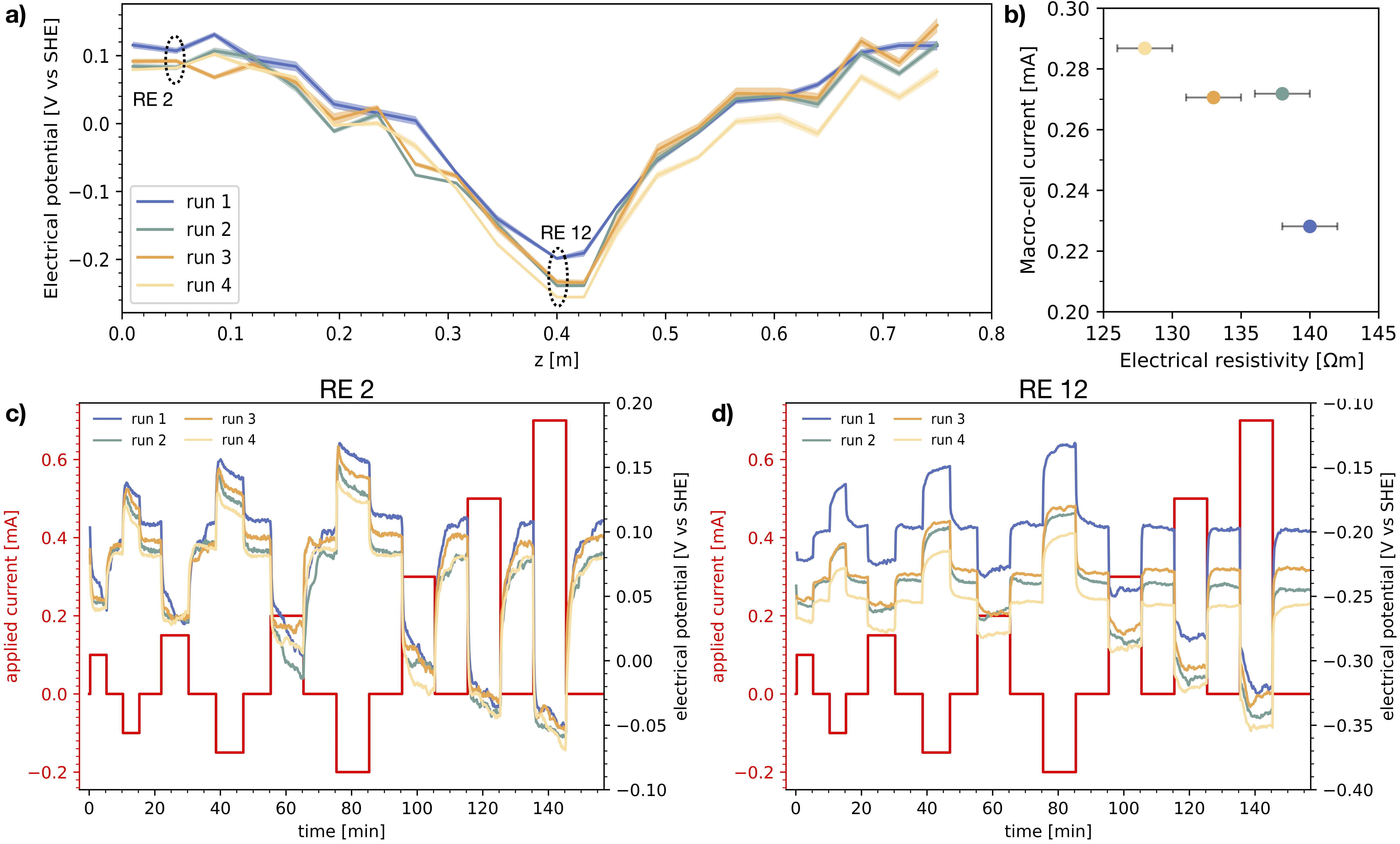}
	\caption{The experimental results of all 4 runs (table \ref{tab:runs}). a) The mean of the OCP, measured when no current is applied directly before each current step (19 measurements for runs 1,2 and 4, 28 measurements for run 3). The shaded regions represent the standard deviation. b) The macro-cell current flowing between the cathode and the anode before the start of the measurement procedure. c) The applied current and measured potential over time, for reference electrode 2, during the measurement procedure with all CE used. d) The same results for reference electrode 12, located directly above the anode.}
	\label{fig:repeatability}
\end{figure}

Figure \ref{fig:repeatability} gives an overview of and compares the potential and macro-cell current measurements obtained during the 4 experimental runs. These results show a good reproducibility of the obtained data in the experimental setup. The OCP over distance (figure \ref{fig:repeatability}a) and the macro-cell current measured between the anode and the cathode (figure \ref{fig:repeatability}b), show only a small variation between the 4 runs. This variation can arise due to factors such as small drift of the RE or the variation in the condition of the steel surface. Part can be explained by the small differences in the electrical resistivity of the solution, as a slight trend of the macro-cell current is visible as a function of the resistivity (figure \ref{fig:repeatability}b), which correlates with the differences in potential directly above the anode at RE 12 (figure \ref{fig:repeatability}a). Figures \ref{fig:repeatability}c and d show the measured potential at RE 2 and RE 12, respectively, during the galvanic pulse procedures (see figure \ref{fig:flowchart}), using CE configuration 'all CE'. Again the variation in potential for the different runs is visible, but these figures also show that the response to the external polarization is very similar in magnitude. \\\\
A further important observation is that the results are stable over time. After applying each current step, the potential generally reached a stable OCP within the assigned time (see table \ref{tab:applied_currents}), before the next pulse is applied (figures \ref{fig:repeatability}c and d). Additionally, figure \ref{fig:repeatability}a shows the mean and standard deviation of the OCP, before each galvanostatic current step during the measurement procedure. The small standard deviation indicates that this OCP is very stable over time. This lets us conclude that the applied currents during the galvanic pulses are small enough, and applied for a sufficiently short duration, so that the corrosion system is not affected by the external polarization in the setup.

\FloatBarrier

\begin{figure}[h]
	\centering
	\includegraphics[width=0.9\textwidth]{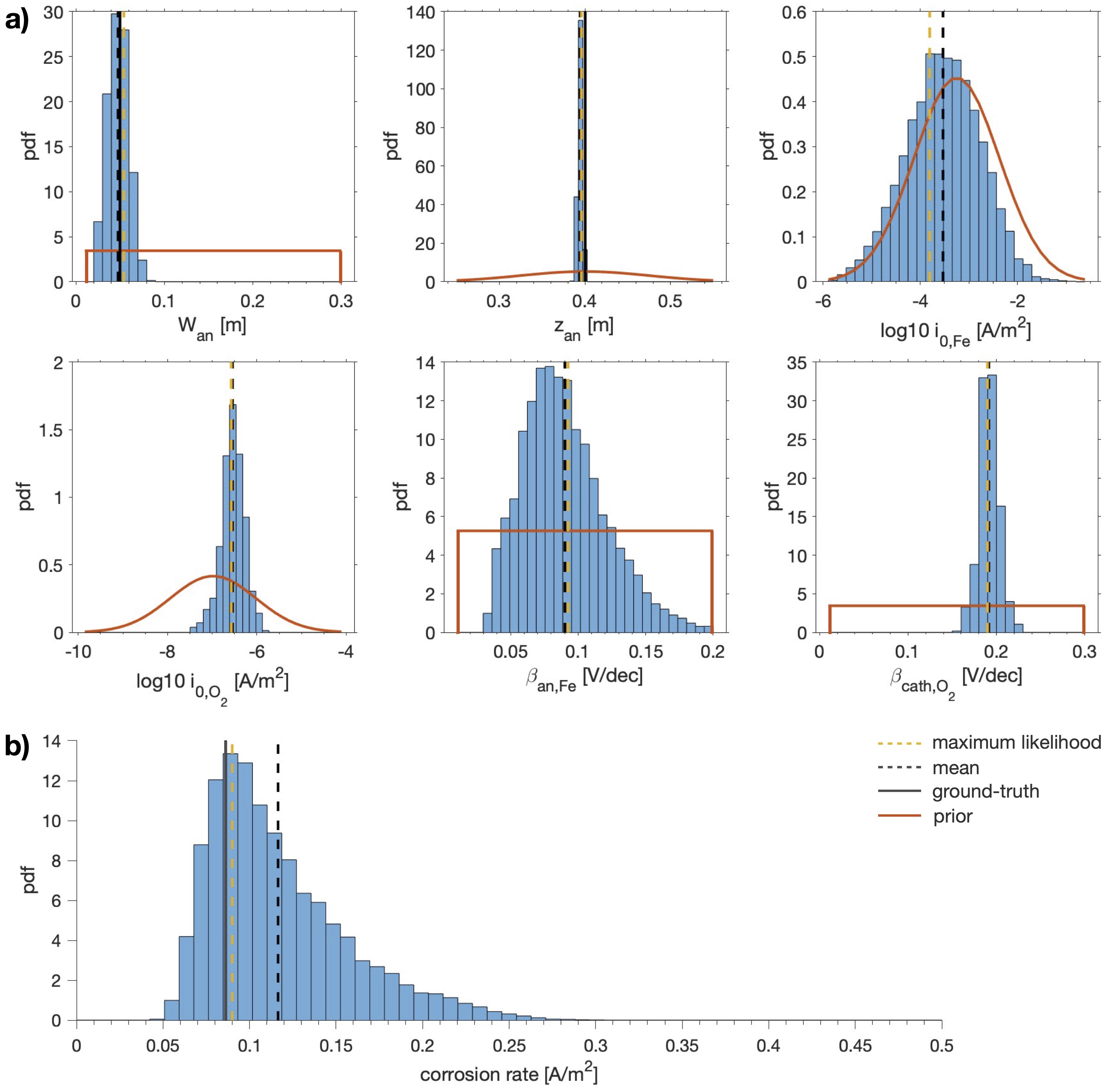}
	\caption{ECT results of experimental run 3, obtained with all data from the three measurements with 9 currents and different CE positions. a) The probability density distributions (pdf, blue) of the 6 model parameters, with the prior probability densities (red). b) The pdf of the corrosion rate. The dashed black lines indicate the mean of the marginal distributions, the solid black lines indicate the ground-truth values for $W_{an}$, $z_{an}$ and the current density at the anode (obtained from the measured macro-cell current between the anode and cathode). The maximum-likelihood models are indicated in yellow.}
	\label{fig:best_results}
\end{figure}

In figure \ref{fig:best_results}, the estimated probability density functions (pdf) are shown for all six model parameters, as well as the corrosion rate, obtained by using ECT for all experimental data of run 3. For this run, the most information was obtained, as it includes the electrical potentials at the surface of the flow-cell, measured for 9 different applied external currents and for three different CE configurations (table \ref{tab:runs}, figure \ref{fig:flowchart}). The results show that ECT is well able to estimate the anode location and size for this experimental setup. In figure \ref{fig:best_results}a, we can see a large gain of information for the width of the anode, $W_{an}$, and the location of the anode, $z_{an}$, when compared to the prior information (indicated in red). The probability distributions are unimodal and symmetric. The distributions can be approximated by a Gaussian, with a standard deviation of 0.01 m for $W_{an}$, and 0.002 m for $z_{an}$. Figure \ref{fig:best_results} also shows the maximum-likelihood model (dashed yellow lines), which are the model parameters that produce the highest posterior probability, and therefore are a function of the prior probability and the fit between the observed data and the numerical model \cite{van2021electrochemical}. For $W_{an}$ and $z_{an}$, this model is similar to the mean of the marginal distributions (dashed black lines) and is a good estimate of the known, ground-truth values of $W_{an}$ = 0.05 m, and $z_{an}$ = 0.4 m (solid black lines).

\FloatBarrier

\begin{figure}[h]
	\centering
	\includegraphics[width=0.4\textwidth]{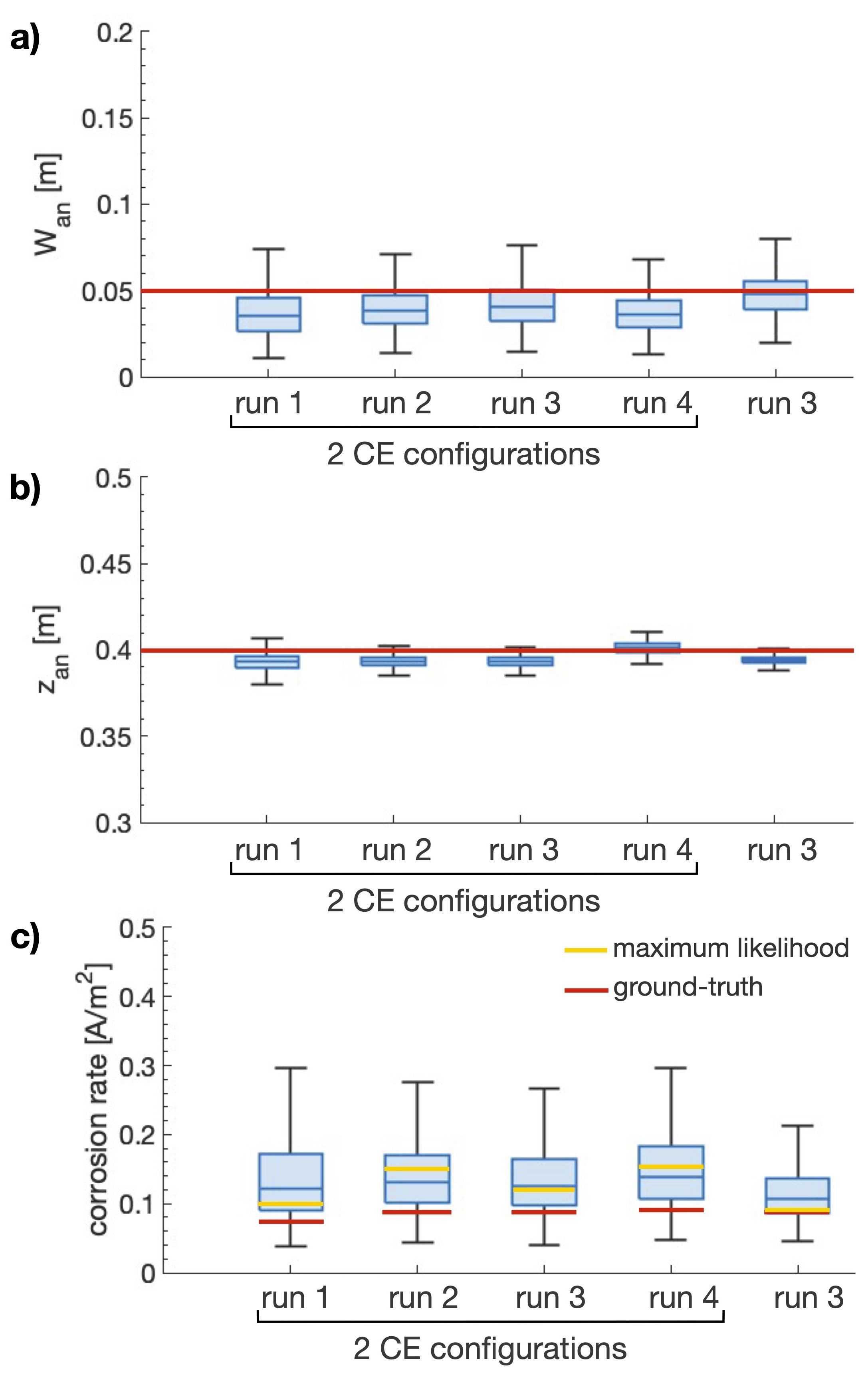}
	\caption{ECT results of all experimental runs obtained with the data of 2 CE configurations ('CE2 \& CE3' and 'all CE'), compared to the results of experimental run 3 obtained with all CE configurations (table \ref{tab:runs}), for a) $W_{an}$, b) $z_{an}$ and c) the corrosion rate, represented as box plots, for all experimental runs. In red the known values for $W_{an}$, $z_{an}$ and the measured current density at the anode (obtained from the macro-cell current between the anode and cathode). In yellow the maximum-likelihood model for the corrosion rates. The outliers are omitted from the box plots. The probability distributions can be found in figure \ref{fig:best_results} and section B of the supplementary materials.}
	\label{fig:boxplot1}
\end{figure}

Experimental runs 1, 2 and 4 only contained 2 CE configurations (table \ref{tab:runs}). However using all data from these configurations, a similar quality of results could be obtained for $W_{an}$ and $z_{an}$. Figure \ref{fig:boxplot1} compares the results for $W_{an}$, $z_{an}$, as well as the corrosion rate, represented as box plots. It also shows the results for run 3, if only the data obtained with 2 CE configurations were used. The probability distributions can be found in section B of the supplementary materials. The representation of a box plot indicates the mean value in the centre of the box, the 25th and 75th percentiles with the bottom and top edges of the box and the extreme values with the whiskers. Figure \ref{fig:boxplot1} shows that all experimental runs are well able to estimate the anode location, within a maximum offset of 0.006 m. Runs 1, 2 and 4, using only 2 CE configurations, have a tendency to estimate a lower $W_{an}$ than run 3 (using all configurations), showing a slight underestimation of 0.012 m maximum. These offsets are well within the accuracy needed in engineering applications. The larger offset of the runs 1,2, and 4 is the result of a smaller amount of information due to missing CE configuration 'CE1 \& CE4', as a similar underestimation is visible for run 3, when only the data of 2 CE configurations are considered. Section \ref{sec:posCE} will discuss this in more detail.

\FloatBarrier

The ECT results for the corrosion rates are shown in figures \ref{fig:best_results}b and \ref{fig:boxplot1}c. It is important to note that this corrosion rate is actually the macro-cell corrosion rate, $i_{macro-cell}$. It will slightly underestimate the total corrosion rate, as $i_{total}$ = $i_{self}$ + $i_{macro-cell}$. ECT neglects $i_{self}$, which is the corrosion associated to local electrochemical cells at the anode, i.e., when an additional cathodic reaction occurs at the anode. Similarly, the experimentally determined corrosion rate in these figures also represents $i_{macro-cell}$, as it is obtained by dividing the measured macro-cell current (measured between the anode and cathode), by the anode surface. The corrosion rate can be well estimated by $i_{macro-cell}$, when $i_{self} << i_{macro-cell}$. The measured $i_{macro-cell}$ in the experimental setup up is around 0.1 A/m$^2$. This high corrosion rate will lead to the depletion of oxygen at the anode, and therefore $i_{self}$ can be assumed to be on the order of corrosion rates measured for uniform corrosion in a neutral de-aerated electrolyte on carbon steel. These rates may be around 0.02-0.03 A/m$^2$ in high conductive solution \cite{vanEdekinetics}, but are found to be even smaller in porous media, such as concrete \cite{stefanoni2018corrosion}. Therefore $i_{self}$ can be expected to be smaller than 10 \% of $i_{macro-cell}$. In the current work, the macro-cell corrosion rate is referred to as the corrosion rate, assuming $i_{self}$ is negligible small in relation to the error range found of ECT (figure \ref{fig:best_results}b).  \\\\
Figure \ref{fig:best_results}b shows that if all CE configurations are considered, the experimentally measured corrosion rate can be well estimated by the maximum likelihood of the probability distribution obtained by ECT for run 3. It is visible that the distribution of the corrosion rate cannot be accurately represented by a Gaussian. Therefore, as opposed to $W_{an}$ and $z_{an}$, the mean values of the marginal distributions and thus the boxplots in figure \ref{fig:boxplot1}c might give a false sense of the overestimation of the corrosion rate by ECT. Instead of comparing the mean to the ground-truth, the maximum likelihood should be used. In figure \ref{fig:boxplot1}c it is visible that if only 2 CE configurations are used, the maximum-likelihood model overestimates the corrosion rate up to a factor of 2. More accurate estimations would be obtained by considering the peak of the marginal distribution of the corrosion rate, which would reduce the overestimation to a maximum of 25 \% (Supplementary materials B). However, if data of CE configuration 'CE1 \& CE4' is included, as is the case for run 3, the offset becomes smaller than 0.004 A/m$^2$, corresponding to an error smaller than 5 \%.  This is substantially better than the error of a factor 2 of the corrosion rate, generally accepted for existing techniques relying on the polarization resistance \cite{andrade1992effect}, and much smaller than the error range obtained with these techniques for localized corrosion \cite{angst2020new}. Moreover, the probability distribution for the corrosion rate also gives a good estimation of the error associated with the solution. 
\begin{figure}[h]
	\centering
	\includegraphics[width=0.8\textwidth]{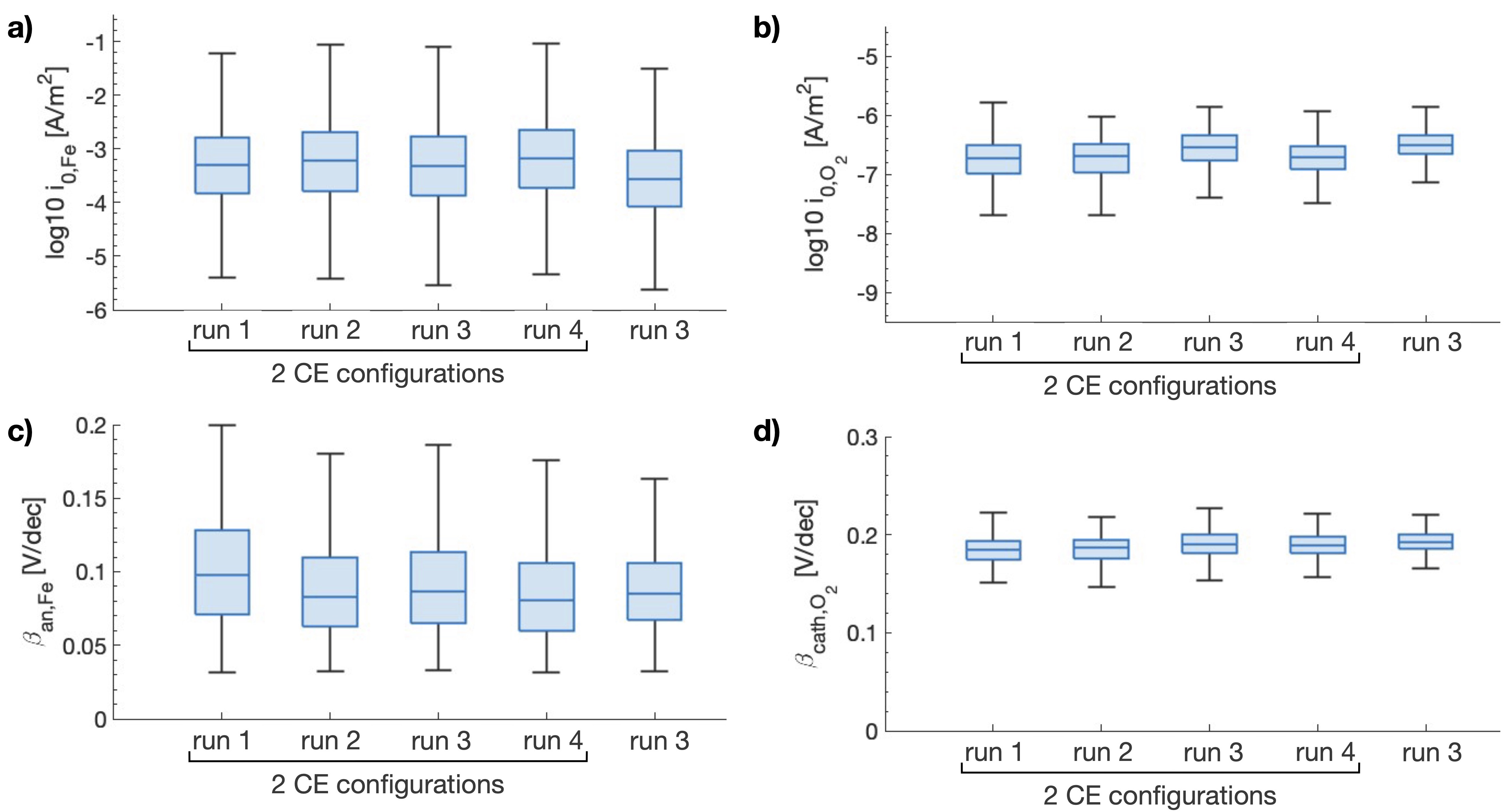}
	\caption{ECT results of all experimental runs obtained with the data of 2 CE configurations ('CE2 \& CE3' and 'all CE'), compared to the results of experimental run 3 obtained with all CE configurations (table \ref{tab:runs}), for a) $i_{0,Fe}$, b) $i_{0,O_{2}}$, c) $\beta_{an,Fe}$ and d) $\beta_{cath, O_{2}}$, represented as box plots, for all experimental runs. The outliers are omitted from the box plots. The probability distributions can be found in figure \ref{fig:best_results} and section B of the supplementary materials.  }
	\label{fig:boxplot2}
\end{figure}
\FloatBarrier
ECT is also able to gain information about the kinetic parameters, the Tafel slopes and exchange current densities. For run 3, the probability distributions obtained by ECT for these parameters show unimodal distributions, with a single peak value close to the mean (figure \ref{fig:best_results}a). Similar results are found for the other three runs (figure \ref{fig:boxplot2}, supplementary materials B). For the ORR kinetics, there is a large information gain compared to the prior. The distributions show a Tafel slope, $\beta_{cath, O_{2}}$, of around 0.2 V/dec, and an exchange current density, $i_{0,O_{2}}$, of around 10$^{-6.5}$ A/m$^2$. These values correspond well to earlier documented values for ORR on stainless steel in a near neutral environment \cite{vanEdekinetics}. The information gain for the anodic iron dissolution reaction kinetics is smaller. The distribution of $i_{0,Fe}$, follows the prior distribution, showing a range around 10$^{-3}$ to 10$^{-4}$ A/m$^2$. The distribution for the anodic Tafel slope, $\beta_{an,Fe}$, shows a peak around 0.08 to 0.11 V/dec. These values are in the range of documented values for iron oxidation in literature \cite{asakura1971electrodissolution, mccafferty2010introduction}.\\\\
In summary, the ECT results show that the method can provide an accurate estimation of the anode size, location and the (macro-cell) corrosion rate. For run 3 all data were included in the inversion that could be obtained in the experimental setup, giving ECT the maximum amount of information as input. It shows that the relatively simple numerical model, based on Butler-Volmer kinetics, the diffusion limited current density, and homogenous conductivity, is accurate enough within the current range of the applied galvanostatic pulses. Furthermore, the amount of data is sufficient to obtain unimodal probability distributions, thus showing a single peak value for all six model parameters. The drawback of using this large amount of input information is that the measurement time to obtain all data, using the current procedure, is around 7.5 hours. Moreover, the computational time of ECT, using 4 cores on the ETH Zurich Euler cluster for the computation of the forward model, takes around three weeks for this amount of data. Decreasing the amount of input data would decrease both the measurement and computation time.\\\\
In the following sections, we first elaborate on the adjustments made to ECT compared to the technique as described in van Ede et al. (2021) \cite{van2021electrochemical}, that were crucial to obtain the presented validation results. Secondly, we discuss two possibilities to decrease the amount of input data needed for ECT, thereby decreasing the measurement and computation time, while still obtaining reliable results. Here, the measurements obtained during run 3 are used, as it contains all three CE configurations. 

\begin{figure}[h]
	\centering
	\includegraphics[width=\textwidth]{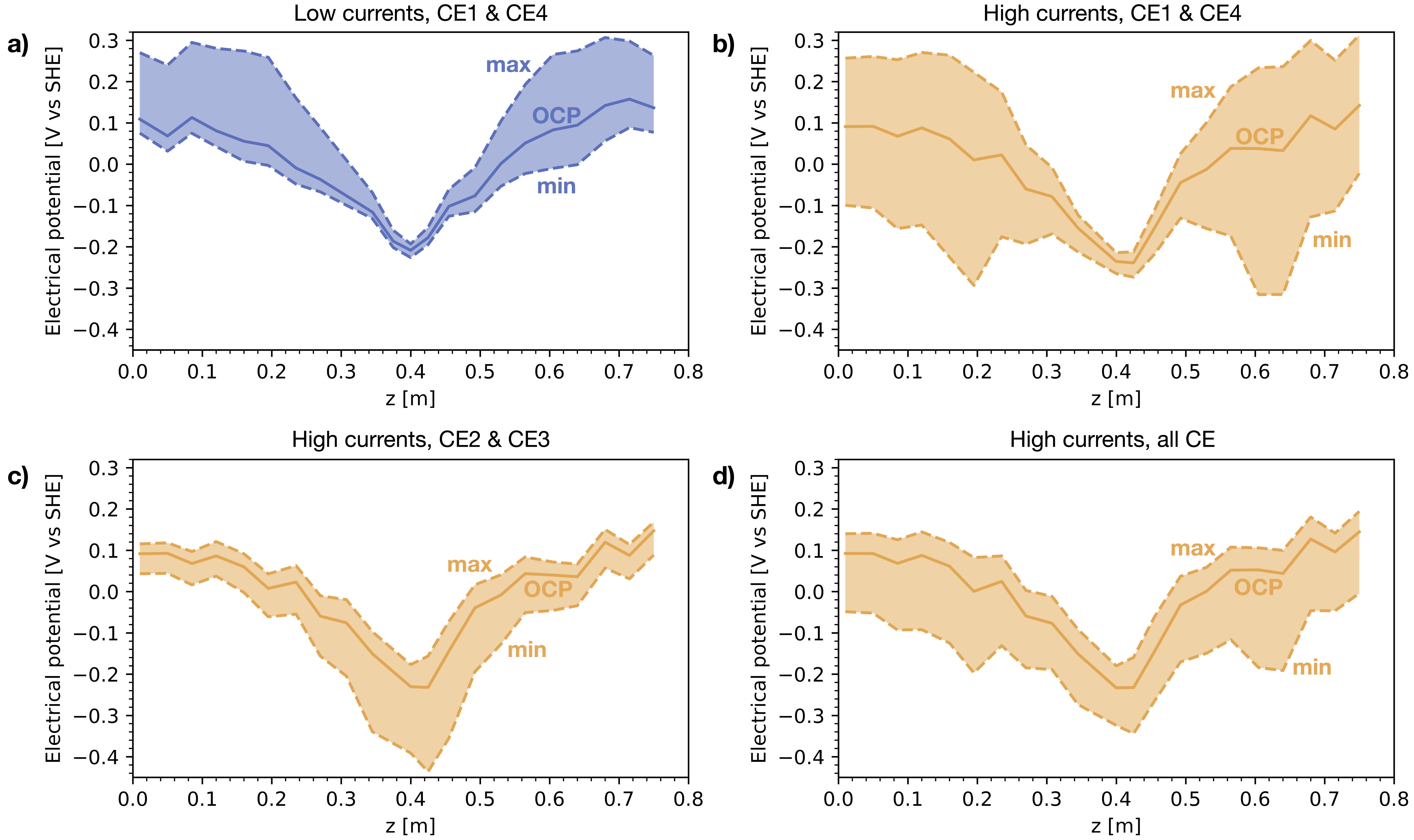}
	\caption{The maximum offset of the potential from OCP to the minimum and maximum applied external current respectively, measured by the RE at the surface of the flow-cell setup. In blue, currents in the range of -0.08 to 0.2 mA, applied  through counter electrodes (CE) in the region of the cathodes of the macro-cell (CE1 \& CE4) (a), as used in van Ede et al (2021) \cite{van2021electrochemical}. In yellow, currents applied in the range of -0.2 to 0.7 mA, applied through CE in the region of the cathodes (CE1 \& CE4) (b), through CE located in the region of the anode (CE2 \& CE3) (c) and using all 4 CE (d), for experimental run 3 (table \ref{tab:runs}).} 
	\label{fig:lowhighcurrent}
\end{figure}
\FloatBarrier

\subsection{Adaptions to the ECT technique}

\subsubsection{Adjusting the current range \&  the positions of the CE} \label{sec:posCE}

ECT as described in van Ede et al. (2021), used a current range between -0.08 and 0.2 mA, applied at CE1 \& CE4 (see figure \ref{fig:exp_overview}), positioned in the region of the cathodes. Figure \ref{fig:lowhighcurrent}a shows the maximum variation in potential measured at the reference electrodes in the flow-cell setup, when applying this current range with CE configuration 'CE1 \& CE4'. It shows that the external polarization predominantly changes the potential at the cathode. However, in the centre at the location of the anode, the stimulation of the system is very small. Thus, in this configuration, most information is obtained about the cathode.\\\\
In the current work we increased the range to a minimum anodic current of -0.2 mA and a maximum cathodic current of +0.7 mA (table \ref{tab:applied_currents}). The presented ECT results (figure \ref{fig:best_results}) showed that our numerical model is accurate within this current range. However, this increase in currents, using only the CE in the regions of the cathode, does not increase the potential variation at the anode (figure \ref{fig:lowhighcurrent}b). To stimulate the anodic region, counter electrodes closer to the anode are needed. This is apparent from the results displayed in figure \ref{fig:lowhighcurrent}. When applying the galvanostatic polarization through CE2 and CE3, positioned closer to the anode, we see a larger potential offset in the region of the anode. Thus, using the additional CE configuration 'CE2 \& CE3', ECT can obtain more information about the anode, compared to the situation when only CE in the regions of the cathode of the macro-cell are used.
\begin{figure}[h]
	\centering
	\includegraphics[width=0.85\textwidth]{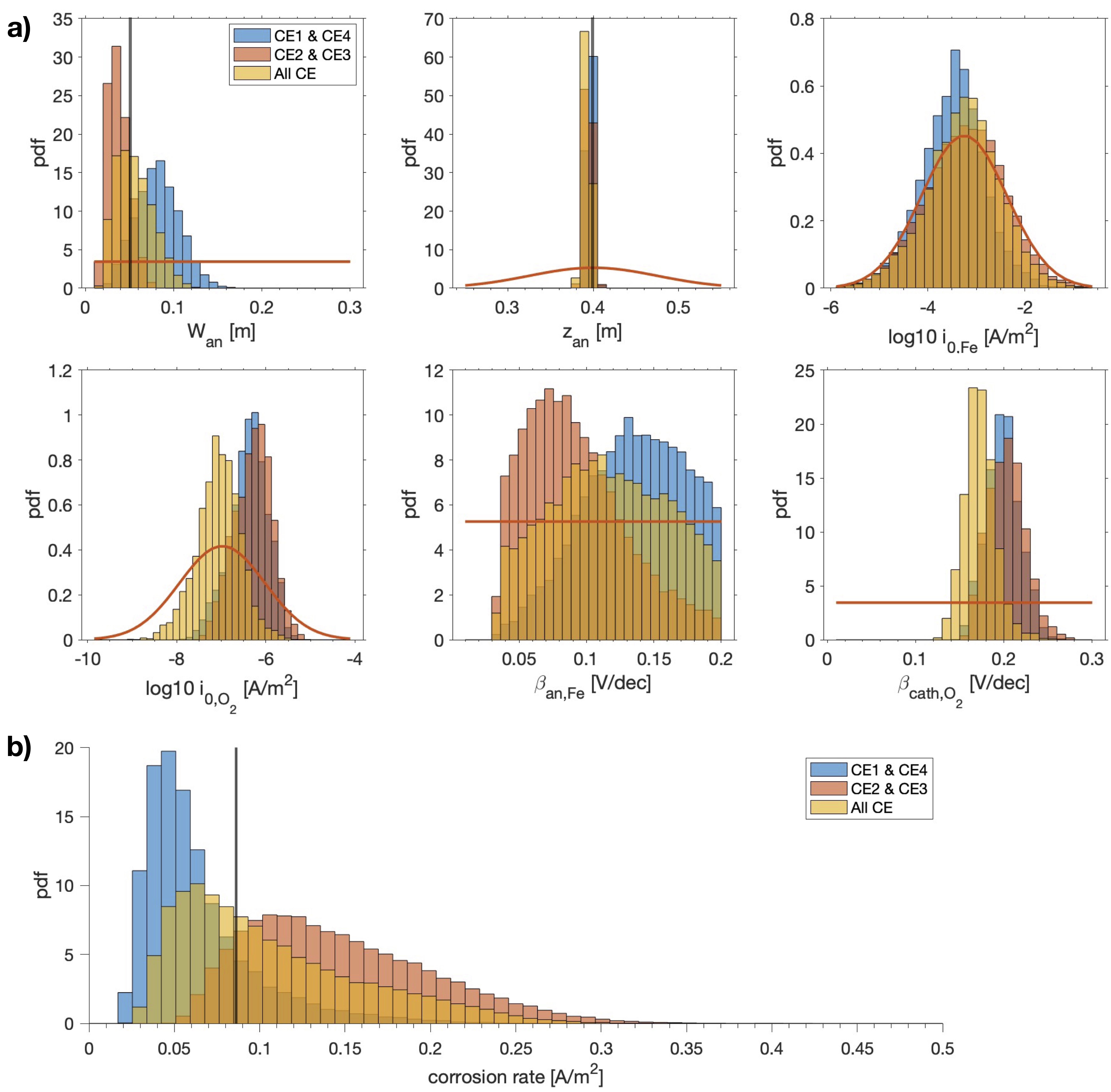}
	\caption{ECT results of experimental run 3, obtained with data of the three different CE configurations: 'CE1 \& CE4' in the region of the cathode, 'CE2 \& CE3' near the anode, and all CE. a) the probability distributions of the model parameters, with the prior probabilities in red. b) the probability distribution of the corrosion rate. The solid black lines, for $W_{an}$, $z_{an}$ indicate the ground-truth values, and the current density at the anode (obtained from the measured macro-cell current between the anode and cathode).}
	\label{fig:positionCE}
\end{figure}
\FloatBarrier
The effect of the use of different CE configurations can also be observed in figure \ref{fig:positionCE}. Here the ECT results are shown, obtained by using the data of the three CE configurations: 'CE1 \& CE4' in the region of the cathode, 'CE2 \& CE3' near the anode, and 'all CE'. If using only the data obtained with the 'CE2 \& CE3' configuration, we underestimate the width of the anode, and overestimate the corrosion rate. However, we obtain more accurate results for the anodic reaction kinetics (closer to the results presented in figure \ref{fig:best_results}). If we only use the 'CE1 \& CE4' configuration, we overestimate the anode size and underestimate the corrosion rate. Using 'all CE' will give us the better estimations of the anode width and the corrosion rate. The most accurate ECT results will be obtained by combining the data off all different CE configurations, as was shown in figure \ref{fig:best_results}. Therefore, leaving out the data for the 'CE1 \& CE4' configuration, results in a slightly poorer estimation for the corrosion location, size and rate as was shown in figure \ref{fig:boxplot1}.
\begin{figure}[h]
	\centering
	\includegraphics[width=0.7\textwidth]{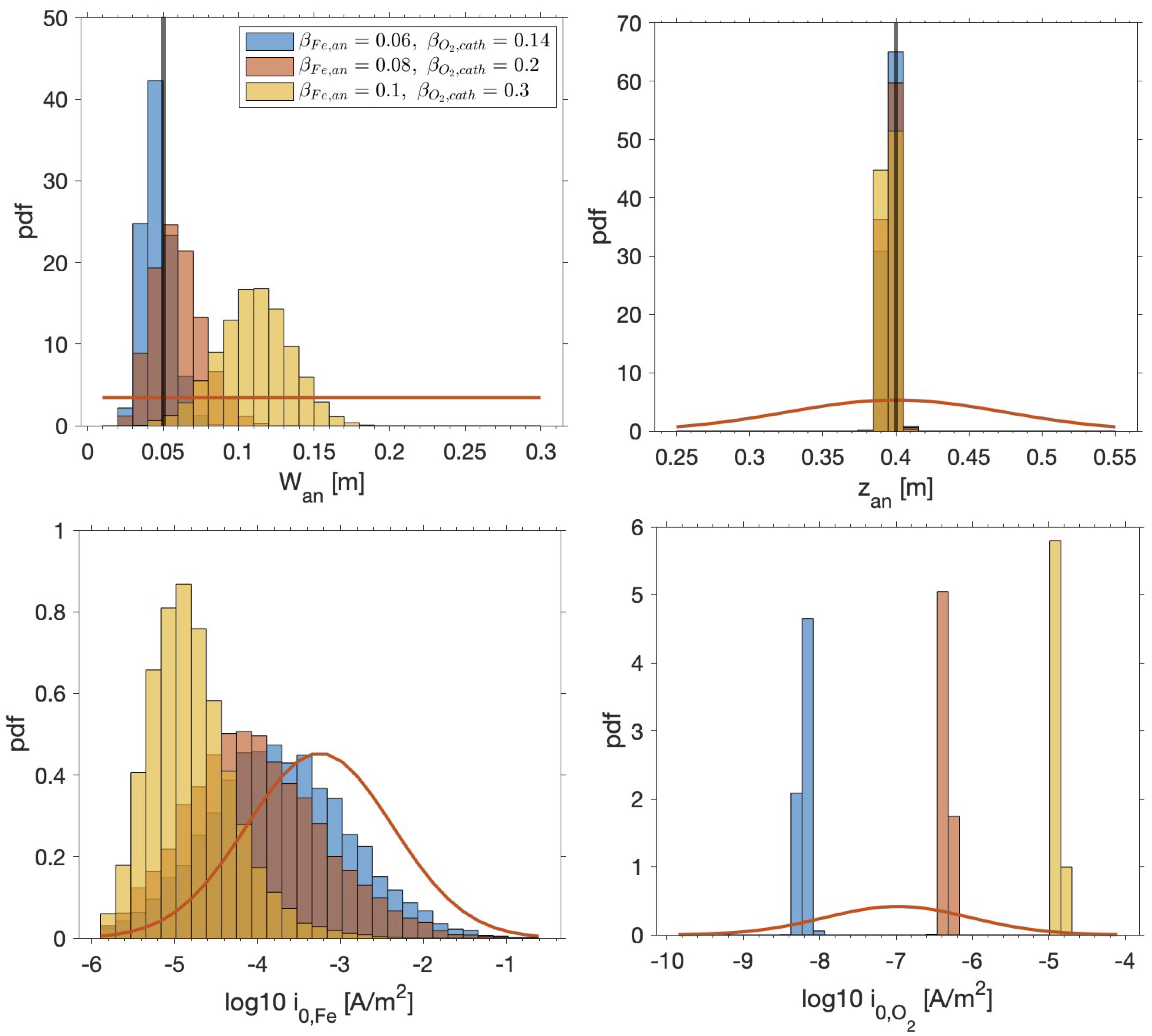}
	\caption{ECT results of experimental run 3, obtained with data of CE configurations 'CE1 \& CE4', for four model parameters, using constant values for the anodic and cathodic Tafel slopes, $\beta_{an, Fe}$ and $\beta_{cath, O_{2}}$. The probability densities (pdf) of the priors are given in red. The solid black lines for $W_{an}$, $z_{an}$ indicate the ground-truth values.}
	\label{fig:4vs6}
\end{figure}
\FloatBarrier

\subsubsection{Optimizing the number of model parameters}

A possible goal of ECT when used as an NDT method, is to estimate the anode size and location, and the corrosion rate of a localized corrosion system. Theoretically, all other parameters in the numerical model may be considered to be constant for a known stable corrosion environment. Here, we want to assess if the number of model parameters (degrees of freedom) could be reduced without compromising the performance of ECT as an NDT method. A first aspect is that accurate data on the kinetic parameters, exchange current densities and Tafel slopes, is scarce. A large spread can be observed for documented values in literature, first because of the subjectivity in the estimation of these parameters from measured potential-current curves \cite{van2022analysis}, and second because of the dependency of these parameters on the measurement methodology \cite{vanEdekinetics}. For this reason, in van Ede et al. (2021) \cite{van2021electrochemical}, the exchange current densities were included as model parameters in the inversion, while the Tafel slopes were considered constants in the numerical model.
\begin{figure}[h]
	\centering
	\includegraphics[width=\textwidth]{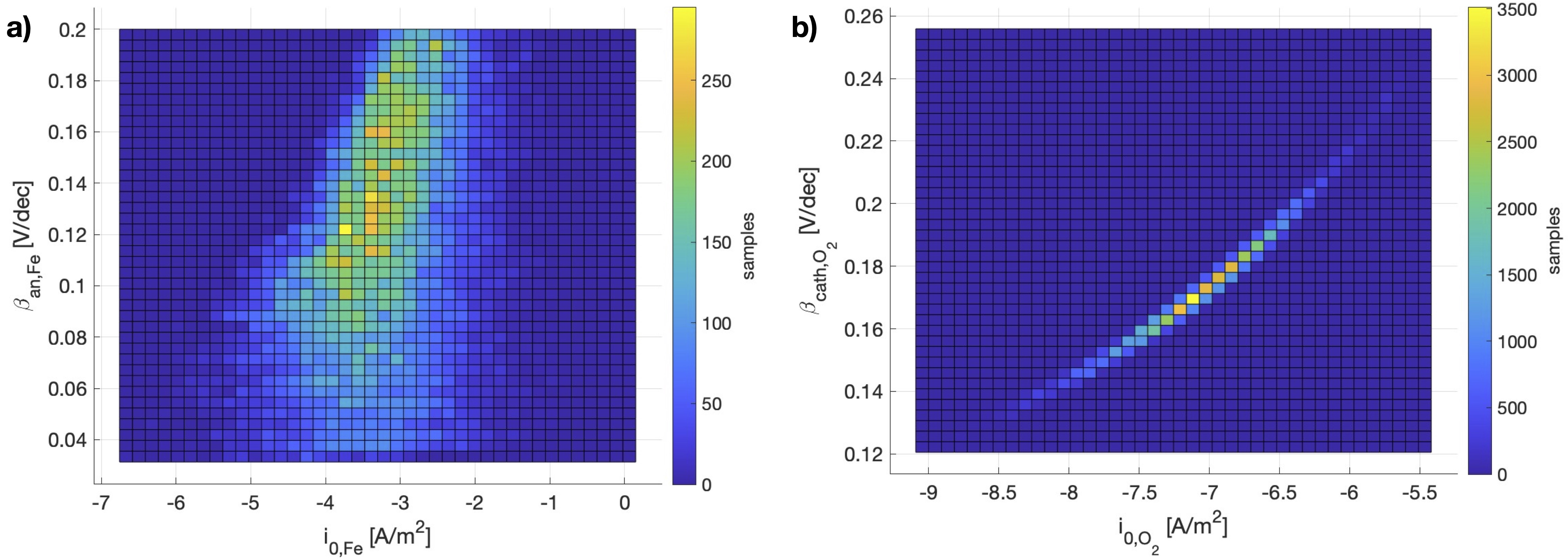}
	\caption{2D histograms of the sampled model parameters for the ECT results of experimental run 3 (figure \ref{fig:best_results}). a) The anodic Tafel slope, $\beta_{an, Fe}$, versus the exchange current density, $i_{0,Fe}$. b) The cathodic Tafel slope, $\beta_{cath, O_{2}}$, versus the exchange current density, $i_{0,O_{2}}$. }
	\label{fig:bvsio}
\end{figure}
\FloatBarrier
In the current study, we observed that fixing the Tafel slopes can lead to a wrong indication of the exchange current densities and the anode size, as is visible in figure \ref{fig:4vs6}. This figure shows the ECT results of experimental run 3, using the data for CE configuration 'CE1 \& CE4' and fixing the anodic and cathodic Tafel slopes, $\beta_{an, Fe}$ and $\beta_{cath, O_{2}}$, to the values indicated in figure \ref{fig:4vs6}. The highest values for the Tafel slopes lead to a clear overestimation of the width of the anode. The probability distributions for the exchange current densities are even more dependent on the chosen values for the Tafel slopes, especially the distribution of $i_{0,O_{2}}$. These results are caused by the trade-off between the Tafel slopes and exchange currents densities, which is visualized in figure \ref{fig:bvsio}. Here, the Tafel slopes are plotted versus the exchange current densities of the anodic and cathodic reaction kinetics in a 2D histogram, showing the number of samples during the MCMC sampling of the probability distribution (section \ref{sec:meth_ECT}). Especially for the cathodic reaction kinetics, it shows that the chosen value for $\beta_{cath, O_{2}}$ directly influenced the probability distribution of $i_{0,O_{2}}$. Therefore, we suggest to include Tafel slopes as model parameters in the inversion of ECT. Figure \ref{fig:best_results} shows that including the different CE configurations, we obtain sufficient information for ECT to be able to solve for all six model parameters.

\subsection{Optimizing ECT}
\begin{figure}[h]
	\centering
	\includegraphics[width=0.9\textwidth]{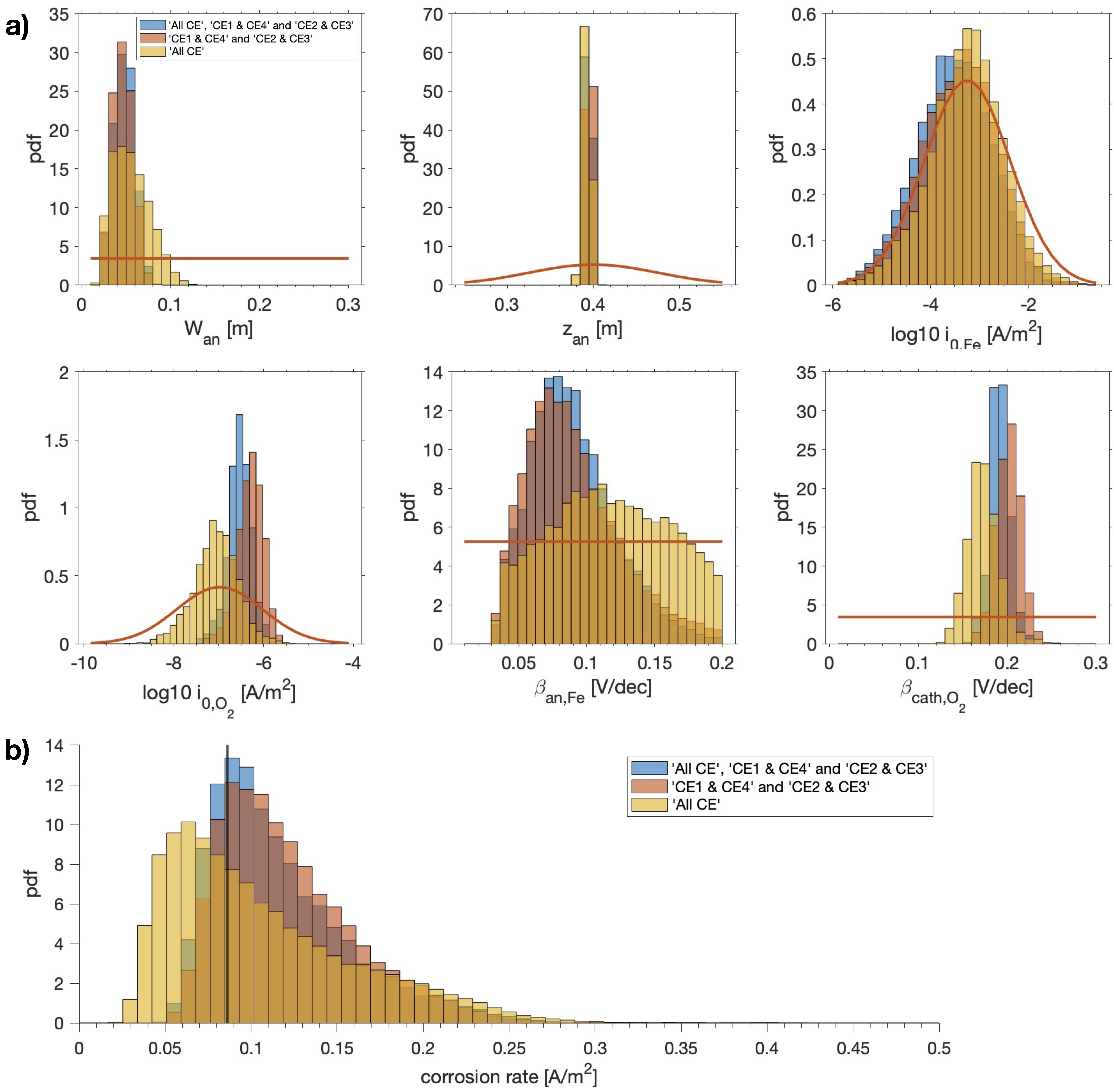}
	\caption{ECT results, comparing the amount of CE configurations used for experimental run 3 (table \ref{tab:runs}). a) The probability density distributions (pdf) of the model parameters, with the prior probabilities in red. b) The pdf of the corrosion rate. The solid black lines, for $W_{an}$ and $z_{an}$ indicate the ground-truth values, and the current density at the anode (obtained from the measured macro-cell current between the anode and cathode).}
	\label{fig:opt_datalenght}
\end{figure}

At this stage, the largest limitation of ECT is the measurement and computation time. This may be largely improved by decreasing the measured input data needed. Figure \ref{fig:opt_datalenght} investigates the number of CE configurations needed to obtain reliable results.  The figure compares the results that were presented in figure \ref{fig:best_results}, for which all three CE configurations were used, to ECT results using only configurations 'CE1 \& CE4' and  'CE2 \& CE3', and results using the configuration 'all CE'. For the former, the configuration 'CE1 \& CE4' should obtain most information for the cathode, while 'CE2 \& CE3' focusses more on the anode (see figure \ref{fig:lowhighcurrent}). 'All CE' on itself should contain information about both the anode and the cathode. 
\FloatBarrier

\begin{figure}[h]
	\centering
	\includegraphics[width=0.9\textwidth]{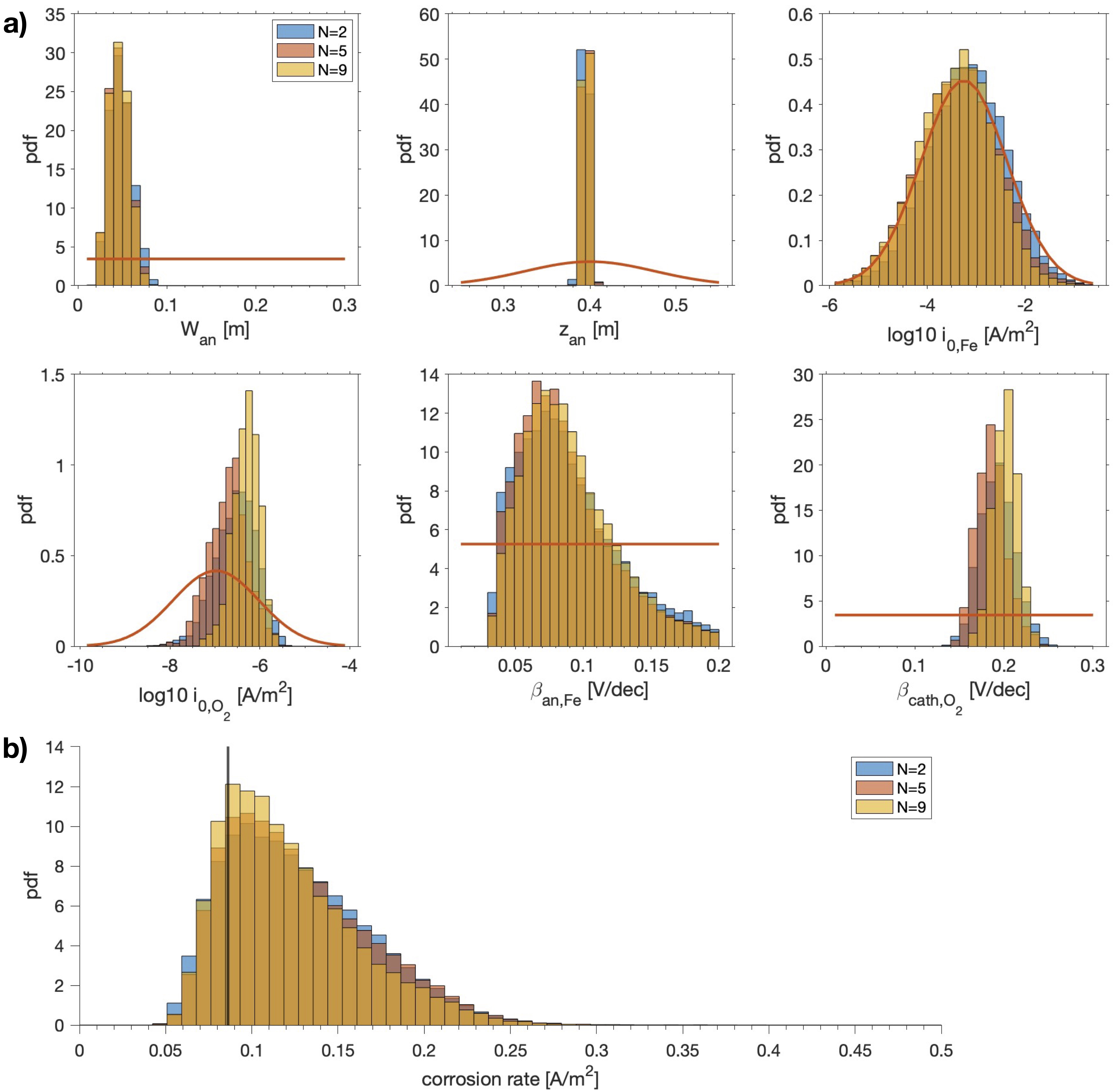}
	\caption{ECT results using data obtained with two CE configurations, 'CE1 \& CE4' and  'CE2 \& CE3' for experimental run 3 (table \ref{tab:runs}), comparing the number of galvanostatic pulses used. N = 9 refers to all pulses listed in table \ref{tab:applied_currents}, N = 5 to [0.2;-0.2;0.3;0.5;0.7] mA, and N = 2 to [-0.2,0.7].  a) The probability density distributions (pdf) of the model parameters, with the prior probabilities in red. b) The pdf of the corrosion rate. The solid black lines, for $W_{an}$ and $z_{an}$ indicate the ground-truth values, and the current density at the anode (obtained from the measured macro-cell current between the anode and cathode).}
	\label{fig:opt_currentlength}
\end{figure}

As was shown in figure \ref{fig:boxplot1}, leaving out configurations 'CE1 \& CE4', reduced the accuracy of the estimations of ECT. However, figure \ref{fig:opt_datalenght} shows that using configurations 'CE1 \& CE4' and  'CE2 \& CE3', would result in probability distributions very similar to the use of all three CE configurations. The need for only 2 configurations would reduce the measurement time by one third. Figure \ref{fig:opt_datalenght} also shows that a single CE configuration ('All CE') is not sufficient to obtain accurate results. \\\\
A further means to decrease both the measurement and computational time, would be to decrease the amount of externally applied currents. Instead of applying 9 different currents (table \ref{tab:applied_currents}), the number of galvanostatic pulses may be decreased to only the 5 highest pulses, or even to only include the maximum anodic and cathodic pulses. This is investigated in figure \ref{fig:opt_currentlength}. From this figure, we can conclude that decreasing the number of galvanostatic pulses to the 2 largest, does not significantly affect the accuracy of the estimation for the anode width, location and corrosion rate. Only the cathodic kinetic parameters are slightly affected. The obtained results show that most information is carried by the highest applied pulses.

\FloatBarrier

\subsection{Outlook}
As highlighted in the introduction, the non-destructive measurement of corrosion rates can be highly beneficial to increase the efficiency of, and thus decrease the costs related to, the corrosion condition assessment and repair. The experimental validation of ECT has shown that it is a promising NDT technique to determine the location, size and rate of localized corrosion. Because it requires the external polarization of the steel, mapping of the corrosion rate over large areas of engineering structures, as is done commonly for the electrical potential, is not realistic. Instead the measurement of the corrosion rate is most valuable in combination with existing commercial techniques, such as half-cell potential mapping. Corrosion rate measurements could be performed at high-risk sites, to gain confidence about the need for repair.\\\\
To apply ECT in the field, further developments should first of all be focussed on decreasing the measurement and computation time. By using only two CE configurations and two externally applied galvanostatic pulses, the measurement time is reduced to 40 minutes, which is still too long for future field applications.  Ideally measurements do not take more than a few minutes, even if it would only be applied on locations with a high suspicion of corrosion. To decrease the measurement time, ideally, the technique should move from steady state measurements, to pulse measurements.\\\\
The reduction of the needed input data would reduce the computational time of the described stochastic inversion approach only by half. This can be reduced by further parallelization of the MCMC sampling, the use of GPUs and/or decreasing the amount of samples to 30,000 - 40,000, as was shown in the convergence assessment in van Ede et al. (2021) \cite{van2021electrochemical}. However, to be able to perform, fast, real-time time corrosion rate measurements in the field, ECT should transition from its described stochastic inverse approach, to a gradient-based deterministic approach. The stochastic inversion allowed us to investigate the inverse problem and optimize for the measured input data needed. It found well-behaved, unimodal probability distributions for all model parameters, of which the maximum-likelihood models were well able to estimate the corrosion location and rate. Instead of the estimation of the complete posterior, deterministic approaches seek to find the maximum-likelihood, by minimizing a misfit function. This misfit function is mainly described by the difference in observed data and the data produced by the numerical model with a set of estimated model parameters. Gradient-based deterministic approaches are computationally fast, but their main limitation is the danger of finding a local minimum of the misfit function, instead of the minimum that corresponds to the maximum-likelihood model \cite{shewchuk1994introduction,fichtner2021lecture}. However, the obtained posterior distributions in the current work have shown that if any initial value is chosen from the prior distributions, the chance of converging towards a local minimum is small. Therefore, the application of deterministic inverse methods in ECT is expected to reliably find the maximum-likelihood model, and could substantially decrease the computation time.\\\\
Finally, the current work showed the validation of a controlled macro-cell corrosion system in solution. Further steps should include the validation in porous media such as soil and concrete. ECT will most likely need further refinements. First of all, in porous media, there are several influences on the electrical potential that are not incorporated in the numerical model, such as diffusion potentials \cite{angst2009diffusion} and a heterogeneous resistivity \cite{rodrigues2021reinforced}. It is to be investigated how much these influences decrease the accuracy of ECT, and if adjustments of the numerical model are necessary. Secondly, the accuracy of ECT will be dependent on the size of the anode. The larger the anode, the smaller the effect of the applied galvanic pulses will be (as shown by the authors in \cite{van2022inverse}), and possibly the currents have to be adjusted accordingly to get sufficient potential responses at the reference electrodes. Finally, the detected anode might consist of multiple smaller anodes. It would be valuable to evaluate the error of ECT in relation to assuming that these form one single anode.  

\section{Conclusions}
The validation presented in the current work has shown that ECT is a reliable non-destructive technique to determine the location, size and rate of localized corrosion. The following major conclusions are drawn:
\begin{enumerate}
	\item The validation of ECT was presented for the study of a localized corrosion system in a controlled laboratory setup, which used a flow-cell to keep the electrolyte constant in terms of aeration and electrical resistivity, and simulated macro-cell corrosion using the galvanic interaction of carbon and stainless steel. Within the current range of the galvanostatic pulses, the numerical model was found to be well able to describe the experimental behaviour. Additionally, the corrosion system was not permanently influenced by the external polarization, which means that the application of ECT can be considered "nondestructive". 
	\item ECT is able to accurately estimate the anode size, location and corrosion rate in the controlled laboratory setup, when the steel was polarized using counter electrodes positioned in the region of the cathode and in the proximity of the anode. The anode size and location were estimated with a standard deviation of less than 1 cm, well within the margin of the accuracy needed in engineering practice. Corrosion rates could be determined within an offset of 5 \% compared to the measured macro-cell currents between the carbon and stainless steel, an accuracy that is much smaller than existing techniques relying on the polarization resistance.
	\item ECT can give us insight into fundamental parameters of macro-cell corrosion. It was able to solve for all kinetic parameters, the Tafel slopes and exchange current densities of the oxygen reduction reaction and iron dissolution, finding well defined and reproducible probability distributions and estimating values well within the range of the literature.
	\item Further optimization of ECT, in terms of the measurement time, was possible by decreasing the number of applied galvanostatic pulses, and the amount of used counter electrode locations. 
\end{enumerate}
Due to our ageing infrastructure, the need for non-destructive testing for the condition assessment of civil infrastructure will keep increasing in the coming years. ECT can be a valuable asset in the early detection and quantification of localized corrosion, allowing for more efficient maintenance, minimizing safety risks and reducing the economic impact of corrosion. 

\section{Acknowledgements}
This work was supported by the Swiss National Science Foundation [projects no. PP00P2\_163675 and PP00P2\_194812].

\section{Data availability}
Raw measurement data of the 4 experimental runs will be made available in an online repository.

\printcredits

\end{document}